\documentclass[useAMS,usenatbib]{mn2e}
\usepackage{graphicx}
\usepackage{color}
\usepackage{bm}

\def\etal{\emph{et al. }}
\def\ie{\emph{i.e. }}
\def\cf{\emph{c.f. }}
\def\apriori{\emph{a priori }}
\def\Apriori{\emph{A priori }}

\title[Direct imaging field of view with diluted apertures]{Direct imaging with highly diluted apertures. I. Field of view limitations}

\author[O. Lardi\`ere, F. Martinache \& F. Patru]{O. Lardi\`ere$^{1}$\thanks{E-mail:
olivier.lardiere@oamp.fr (OL); frantz@astro.cornell.edu (FM);
fabien.patru@obs-azur.fr
(FP)}, F. Martinache$^{2}$ \& F. Patru$^{3}$\\
$^{1}$Coll\`ege de France, Observatoire de Haute-Provence, Saint-Michel-l'Observatoire, F-04870, France\\
$^{2}$Department of Astronomy, Cornell University, Ithaca, NY
14853, USA\\
$^{3}$Observatoire de la C\^ote d'Azur, Dpt. Gemini, UMR CNRS
6203, avenue Copernic, Grasse, F-06130, France}
\begin{document}

\date{Accepted 2006 November 30. Received 2006 November 21; in
original form 2006 March 7}

\pagerange{\pageref{firstpage}--\pageref{lastpage}} \pubyear{2006}

\maketitle

\label{firstpage}

\begin{abstract}

Future optical interferometric instrumentation mainly relies on
the availability of an efficient cophasing system: once available,
what has so far postponed the relevance of direct imaging with an
interferometer will vanish. This paper focuses on the actual
limits of snapshot imaging, inherent to the use of a sparse
aperture: the number of telescopes and the geometry of the array
impose the maximum extent of the field of view and the complexity
of the sources. A second limitation may arise from the beam
combination scheme. Comparing already available solutions, we show
that the so called hypertelescope mode (or densified pupil) is
ideal. By adjusting the direct imaging field of view to the useful
field of view offered by the array, the hypertelescope makes an
optimal use of the collected photons. It optimizes signal to noise
ratio, drastically improves the luminosity of images and makes the
interferometer compatible with coronagraphy, without inducing any
loss of useful field of view.

\end{abstract}



\section{Introduction}

Over the last three decades, important scientific results have
been obtained from long-baseline optical and infrared stellar
interferometers concerning the stars and their environment, and
more recently extragalactic sources \citep{agn_vlti}. These
results have been obtained thanks to sophisticated observing
techniques, such as fringe visibilities and closure phase
measurements \citep{baldwin86}.

Current interferometers involve between 2 and 4 telescopes only.
The study of very complex and/or faint sources therefore requires
many observations and image reconstruction techniques like
aperture synthesis. Future interferometers should involve a large
number of apertures ($> 10$) in coherence (ideally in phase), but
fringe visibility measurement appears no more suitable for such
rich arrays. Indeed, visibility and closure phase measurements
require either a pair-wise combination on different detectors, or
an all-to-one combination in a non-redundant configuration, in
order to isolate the signal provided by each baseline. This is a
constraint that is difficult to satisfy in practice with a large
number of beams in wide band. Moreover these combination schemes
are generally not compatible with stellar coronagraphy, for which
a direct image featuring a bright central interference peak is
required \citep{labey96}. For all these reasons, direct imaging
involving an all-to-one beam combiner seems to be an elegant and
simpler way to exploit a well-populated optical or infrared array.

However, many questions remain concerning the actual performances
of future large arrays devoted to direct imaging, such as the
field of view (FOV), the dynamic-range and the sensitivity. These
parameters are crucial because they will impose the top-level
requirements for the concept of future large interferometers,
according to the expected science cases. This paper aims at giving
some answers about the FOV of interferometers.

Owing to emerging Extremely Large Telescope projects, future
optical interferometric arrays should exhibit very long baselines,
typically kilometric, in order to really offer a complementary
observing approach in terms of angular resolution. In this
context, we consider here only highly diluted arrays. The
associated pupil filling rate, given by:

\begin{equation}
r = n_T \times \frac{d^2}{D^2}\:\:,
\end{equation}

\noindent tends towards zero, with $n_T$ the number of telescopes
of the array, $d$ the diameter of an elementary aperture and $D$
the diameter of the whole interferometer. With a well populated,
diluted array, for which we are only interested in high spatial
frequencies and not by those measured by one elementary aperture
(which therefore excludes LBT kind interferometers
\citep{refLBT}), direct imaging has been proved feasible thanks to
Labeyrie's pupil densification technique
\citep{labey96,pedretti,gillet}.

In his paper, Labeyrie tells us that direct imaging at the focus
of a diluted array is possible if one densifies the pupil, either
by zooming each elementary aperture or by moving them closer to
each other, with a significant gain in sensitivity. The only
condition is to keep the geometry of the array intact.

\Apriori, as long as one does not mix the frequencies sampled by
the interferometer, the remapping of the pupil proposed by
Labeyrie neither adds nor removes any useful information. However,
being always compared to the purely homothetic (Fizeau)
combination scheme, the so-called hypertelescope is known to
provide direct imaging indeed, but on a limited FOV only. The
notion of FOV for an interferometer is somewhat delicate, and
actually requires the introduction of four different FOV. This
distinction is essential to demonstrate that, the hypertelescope
is an optimal optical image reconstruction technique inducing no
useful FOV loss at all.

Section \ref{sec:archi} highlights the influence of the array
geometry on the FOV. Section \ref{sec:bcs} presents beam
combination schemes already available for direct imaging whose FOV
properties are compared in section \ref{sec:comp} and discussed in
section \ref{sec:dis}.

\section{Influence of the array configuration on the FOV}
\label{sec:archi}

\begin{figure*}
\hfill\includegraphics[width=.95\textwidth]{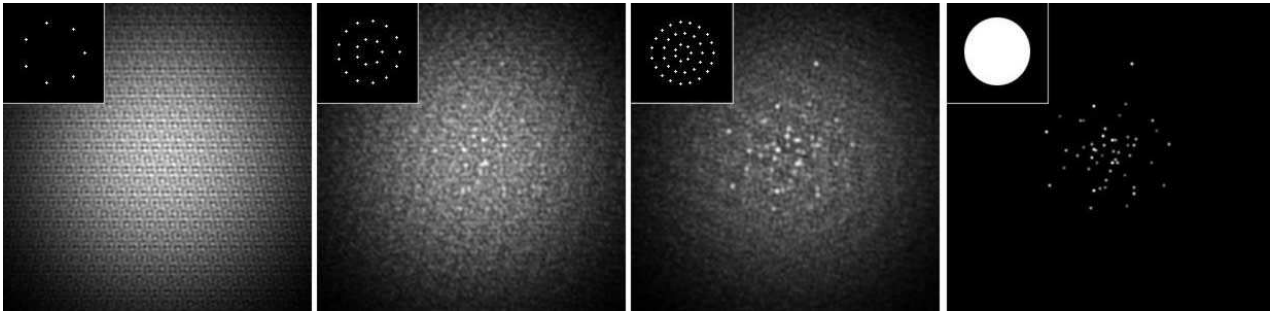}
\hspace*{\fill} \caption{Direct images of a 50-stars cluster
provided by a diluted interferometer involving 7, 20 and 39
apertures compared to a full-filled aperture (right). The crowding
limit of each interferometer is respectively 5, 44 and 169 sources
for $SNR=3$. The 7-aperture case (1$^{st}$ image) is well over the
limit, but the 20-aperture case (2$^{nd}$ image), where only the
brightest stars exceed the background, corresponds to the crowding
limit.} \label{fig:crowding}
\end{figure*}

\subsection{Number of apertures}
\label{sec:nt}

Most current optical interferometers involve less than 4
telescopes working simultaneously. The observables, \ie
visibilities and closure quantities, are used in an inverse
approach to constrain a model of the observed source: uniform or
limb-darkened disk, binary system, etc. They are rarely sufficient
to reconstruct a map or an image of the source, unlike what is
achieved in radio interferometry (a recent example:
\citet{gitti06}) and with aperture masking \citep{tuthi99}. Little
therefore has been written about the limits of wide field imaging
capabilities of an optical interferometer.

An interesting approach by \citet{koech03} and \citet{koech03spie}
uses Shannon's theory of information to give a limit to the
field-resolution ratio. The maximal amount of information that an
interferometer can provide is proportional to the square root of
the number of unique baselines in the array. In case of an
extended filled source, this information amount can be converted
in field-resolution ratio. For a non-redundant interferometer,
this ratio is \citep{koech03}:

\begin{equation}
\mathrm{FOV} / \mathrm{resolution} \leq \sqrt{n_T\times(n_T-1)},
\end{equation}

\noindent with $n_T$ the number of apertures in the array. This
relation gives an upper limit to the usable FOV. We shall call
this limit Information Field (IF). The IF only depends on the
number of apertures and the geometry of the array and not on the
choice of a beam combiner scheme.

One can now reformulate the IF limitation another way. Because of
the finite number of sub-apertures, which only offers a partial
coverage of the spatial frequencies plane, an interferometer
cannot provide the image of an arbitrarily complex extended
source: this is known as the \emph{crowding limit}. Let us
decompose an extended source as a sum made of $p$ elementary (\ie
non-resolved) sources. A point-like source seen by the
interferometer can give a central peak surrounded by a halo or
just a halo (Fig. \ref{fig:psf-off-axis}).

The intensity of each central peak, resulting from the sum of
$n_T$ coherent contributions, is proportional to $n_T^2$. The
average intensity of the halo however, is proportional to $n_T$,
like its RMS fluctuation. After adding-up $p$ elementary
point-like sources, the resulting RMS fluctuation of the halo is
now proportional to $\sqrt{p}\times n_T$.

Yet, a peak remains detectable by (incoherent) substraction of the
halo if it dominates the fluctuation of this halo. With $SNR$
representing the signal-to-noise ratio that one desires the
detection to have, this condition can be written as $n_T^2 \geq
SNR \times \sqrt{p}\times n_T$, which imposes the crowding limit:
\begin{equation}
p \leq n_T^2/SNR^2,
\end{equation}

\noindent meaning that the number of observed sources must be less
than the square of the number of apertures (Fig.
\ref{fig:crowding}). This crowding limit is of course at one with
the field limit that was highlighted earlier: an image made of
$n_T \times n_T$ resels (resolution elements) cannot obviously
provide information on more than $n_T^2$ elementary sources. The
IF can therefore be redefined as the maximal angular diameter of
an extended filled source that can be directly imaged with
$SNR=1$.

\subsection{Geometry of the array}
\label{sec:geo}

\begin{figure*}
\hfill\includegraphics[width=.9\textwidth]{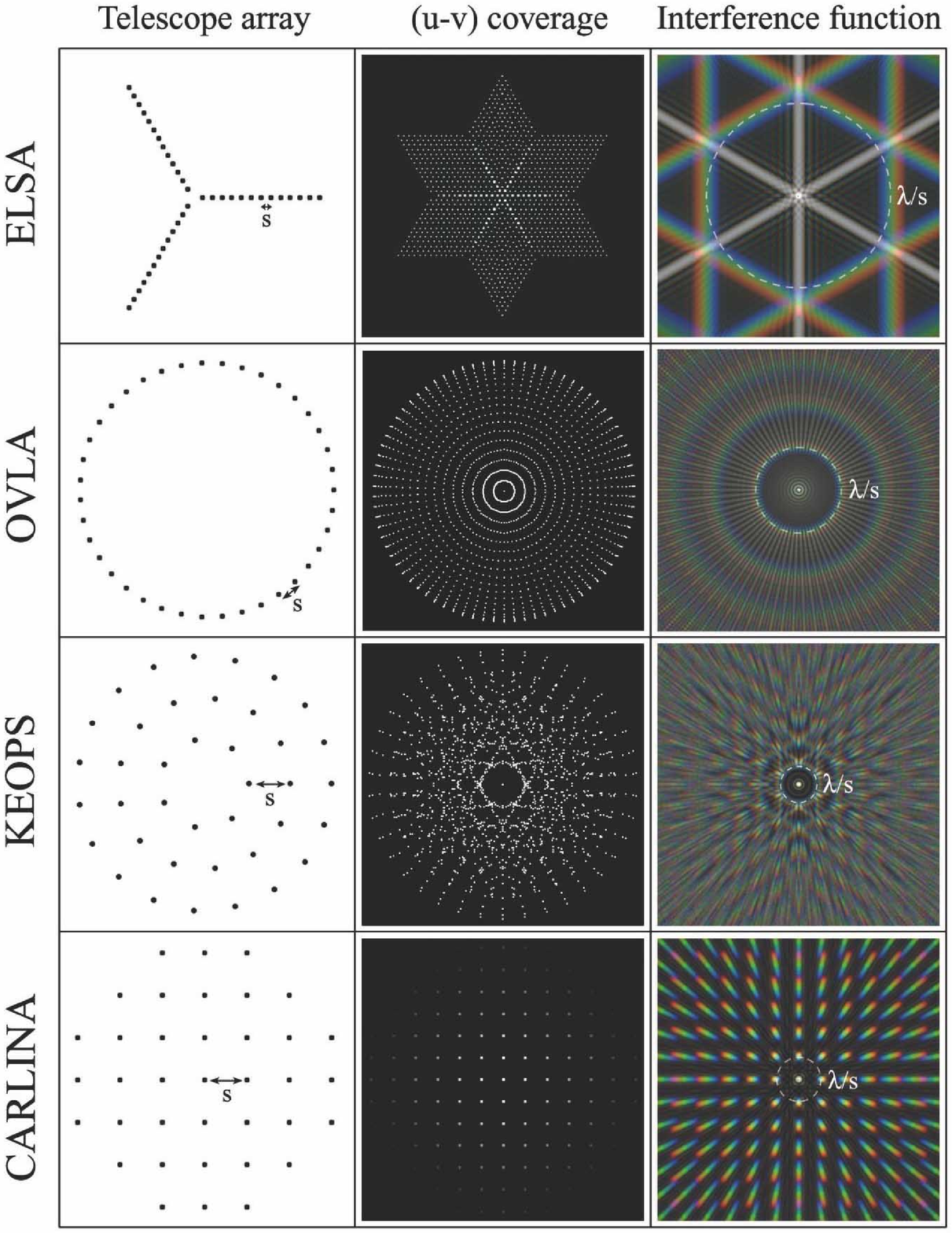}
\hspace*{\fill} \caption{Telescope array configurations of four
different interferometer proposals with their corresponding
$(u,v)$-coverages (in grey levels) and their interference
functions, \ie PSF without envelope (polychromatic images with
$\Delta\lambda/\lambda=0.2$, intensity scale in power $0.3$). For
a fair comparison, these arrays all have the same maximum baseline
and involve 39 telescopes (37 for CARLINA). The radius of the
clean part of interference functions (\ie the CLF size) is
$\lambda/s$, with $\lambda$ the central wavelength and $s$ the
typical minimum spacing between telescopes.} \label{fig:geometry}
\end{figure*}

\begin{figure*}
\hfill\includegraphics[width=.95\textwidth]{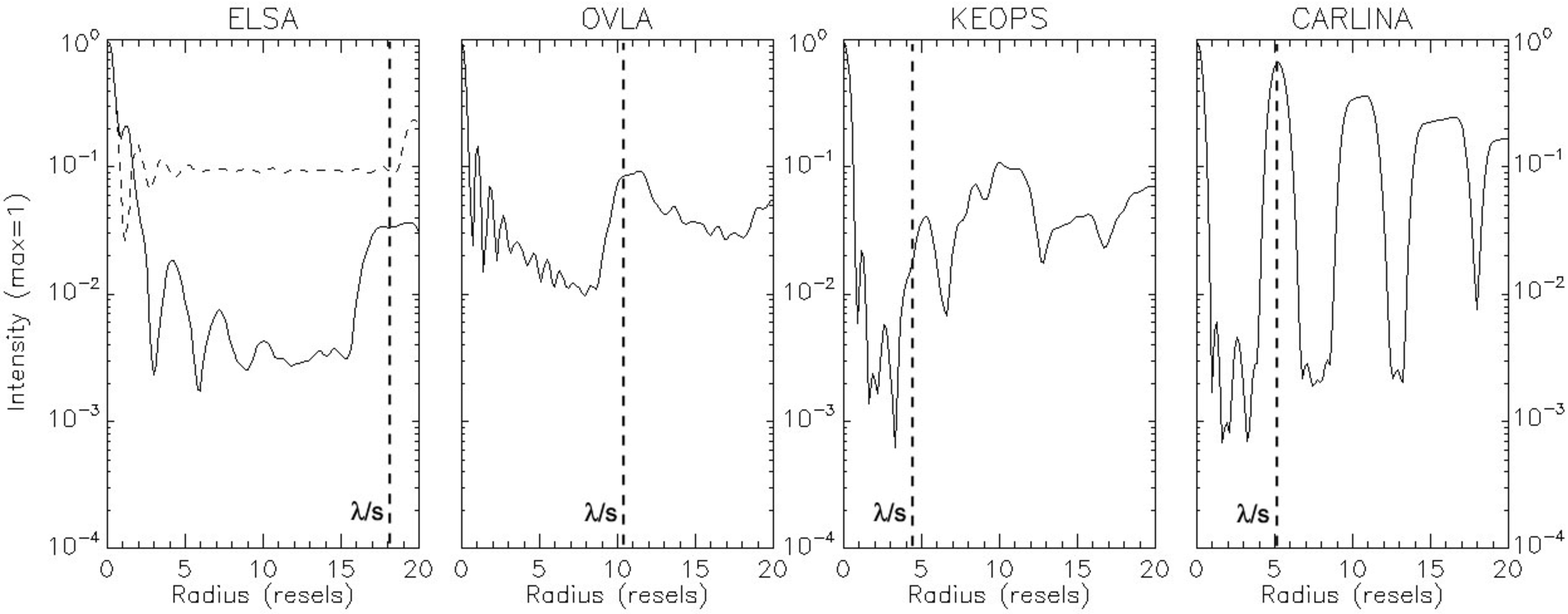}
\hspace*{\fill} \caption{Radial profiles of the interference
functions shown in figure \ref{fig:geometry} (maximum intensity
normalized to 1, radius expressed in resels, with $1\,
\mathrm{resel}=\!\!1.22\,\lambda/D$). The best contrast of the
image inside the Clean FOV ($\lambda/s$) is from $10^{-1}$ (in the
vertical direction: dashed curve) to $3\cdot10^{-3}$ (in the
horizontal direction: solid curve) for ELSA, $10^{-2}$ for OVLA,
$2\cdot10^{-3}$ for KEOPS, and $8\cdot10^{-4}$ for CARLINA.
Uniform arrays, as KEOPS and CARLINA, have a better dynamic-range,
but a narrower CLF, than ELSA and OVLA.} \label{fig:profile}
\end{figure*}

The number of sub-apertures imposes the ultimate limits of FOV and
crowding. Yet, the geometry of the array imposes the practical
limits: the number of unique baselines and the PSF quality.

As with any classical telescope, the PSF of an interferometer is
given by the Power Spectrum of the wavefront complex amplitude in
the output pupil plane \citep{goodman96}. For $n_T$ identical
sub-apertures of diameter $d$ and position vector $\bm{\rho_i}$,
the PSF is given by:
\begin{equation}
I(\bm{\alpha}) = A(\bm{\alpha}) \times \bigg|\sum_{i=1}^{n_T} \exp
\frac{-2i\pi\: \bm{\alpha} \cdot \bm{\rho _{i}}}{\lambda}\bigg|^2,
\label{equ:gen0}
\end{equation}

\noindent with $\bm{\alpha}$ the position vector in the image
plane. Equation \ref{equ:gen0} reminds that this PSF is nothing
but the product of two terms: an interference function, given by
the exponential sum that is determined by the (eventually
remapped) geometry of the array and an envelope $A$, whose shape
and position depend on the retained beam combiner (Sec.
\ref{sec:bcs}).

For a well-populated array of maximum baseline D, the interference
function is very similar to the diffraction pattern of a
fully-filled aperture of diameter D (central peak and Airy rings),
surrounded by a halo of sidelobes (speckles or high-order
dispersed peaks) due to the holes of the pupil plane. According to
the sampling theorem, the angular radius of the clean central part
of the PSF is defined by $\lambda/s$, with $s$ the distance
between two adjacent apertures. For a non-regular array pattern,
the clean part exhibits smoother edges, but we assume that
$\lambda/s$ still gives a good estimate of its mean radius, with
$s$ the ``typical'' smallest baseline in the array.

Consequently, two point-like sources can be observed properly and
simultaneously only if their angular separation is lower than
$\lambda/s$. Then, we can introduce the notion of a clean FOV
(CLF) whose extent is simply:

\begin{equation}
\mathrm{CLF}=\lambda/s. \label{equ:clf}
\end{equation}

One can demonstrate that the CLF is always smaller than the IF,
whatever the array configuration. Indeed, expressed in units of
$\lambda/D$, the diameter of the CLF and of the IF are
respectively $D/s$ and $n_T$. Then the condition $\mathrm{CLF}
\leq \mathrm{IF}$ implies $D \leq n_T \times s$, which is always
true for a 2-dimensional array (there is equality for a linear
regular array). A CLF smaller than the IF means that the crowding
limit is not an issue, provided that one observes a source smaller
than the CLF.

To illustrate the notion of CLF, figure \ref{fig:geometry}
compares the interference function (\ie the PSF without the
envelope) of four different array configurations involving 39
apertures diluted over the same maximum baseline. These
configurations actually correspond, with more or less fidelity to
current proposals for future large interferometric arrays, whose
names will therefore be used for convenience:

\begin{itemize}
\item {\bf ELSA} \citep{elsa} is made of 13 regularly spaced
telescopes along the 3 identical arms of a Y-configuration. ELSA
can, to some extent, be considered as an anti-spider structure
that traditionally bears the secondary mirror of telescopes.

\item {\bf OVLA} \citep{ovla}, whose 39 telescopes are arranged
non redundantly along a circle, can be considered as a giant
telescope with a very important central obscuration.

\item {\bf KEOPS} \citep{keops}, whose telescopes are arranged non
redundantly along three concentric rings of 9, 13 and 19
telescopes, may be compared to a non-obstructed aperture.

\item {\bf CARLINA} \citep{corol04}. Even if the geometry of the
array is not fixed yet, a completely redundant square grid is
often considered. The configuration retained here only uses 37
telescopes to provide a well-balanced array, inside a circle.
\end{itemize}

A glance at figure \ref{fig:geometry} makes us identify two
distinct sparse aperture families. On the one hand, we have the
OVLA and ELSA arrays, which definitely give priority to a
relatively dense and homogeneous coverage of the $(u,v)$ plane.
They exhibit an ``in-line'' geometry, with therefore little space
between telescopes and a rather extended CLF, whose diameter
increases in proportion to the number of telescopes. ELSA presents
preferred axis along which there is redundancy, whereas OVLA's
$(u,v)$ coverage can be described by a purely radial function.
This, of course, results in differences in the interference
function: a centro-symmetric clean field for OVLA and the
appearance of diffraction spikes for ELSA. For a better
comparison, interference function profiles are sketched at figure
\ref{fig:profile}.

On the other hand, there are the KEOPS and CARLINA arrays for
which the priority is to have a uniform coverage of the pupil
plane itself. The distance separating telescopes is therefore
naturally enlarged which induces a reduction of the CLF: its
diameter now only increases in proportion to the square root of
the number of telescopes.

From the strong reduction of the CLF in the case of uniform arrays
such as KEOPS and CARLINA, one may be tempted to exclude those
configurations for direct imaging. However, what is lost in field
is gained in dynamic-range: KEOPS and CARLINA offer a narrower but
darker CLF than ELSA and OVLA (Fig. \ref{fig:profile}).

People working in stellar coronagraphy know that a mandatory
condition to reach very high dynamic-range is at least a telescope
with no central obscuration, possibly optimized by a prolate
spheroidal apodization \citep{soum02}. In those conditions, far
from being uniformly flat, the associated Modulation Transfer
Function or MTF (\ie the $(u,v)$ plan) exhibits a somewhat
``conic'' shape. This requirement is no different for the geometry
of a diluted interferometer if it is made for high contrast
imaging: the coverage of the pupil plane has to be privileged at
the expense of the coverage of the $(u,v)$ plane (Fig.
\ref{fig:geometry}). This statement concurs with the conclusions
of \citet{aime03} who find that the best parameter to evaluate the
relevance of the geometry of a diluted array designed for
exoplanet detection is the integral of the square modulus of its
MTF.

The choice of the geometry of the array must therefore be
motivated by the primary scientific goal of the interferometer.
OVLA and ELSA are definitely made for imaging of wide fields
(multiple or extended sources such as interacting binary stars,
resolved stellar surfaces, envelopes and disks). KEOPS and CARLINA
are better suited for high contrast imaging of compact sources and
for exoplanet detection.

\section{Beam combination schemes}
\label{sec:bcs}

This section focuses on the beam combination scheme, which
determines the shape and the extent of the fringe envelope $A$
(Equ. \ref{equ:gen0}). The beam combiner can limit the FOV
provided by the array if envelope $A$ is narrower than the CLF.
One has to introduce another FOV limitation, that does not depend
on the geometry of the array but only on the beam combination
scheme. This field is referred to as Direct Imaging FOV (DIF), \ie
the FOV inside which an image of a source can be formed directly.
By definition, a point-like source is located inside the DIF if
the central interference peak remains within envelope $A$.

Until now, three beam combination schemes have been considered for
direct imaging: the Fizeau scheme, the Densified Pupil scheme
\citep{labey96}, the IRAN scheme \citep{vakili04} and their fibred
versions \citep{patru06}. This section introduces a common
formalism that describes these combiners and allows an homogeneous
and quantitative comparison of their FOV properties.

\subsection{General formalism}
\label{sec:bcs:form}

\begin{figure*}
\hfill\includegraphics[width=.8\textwidth] {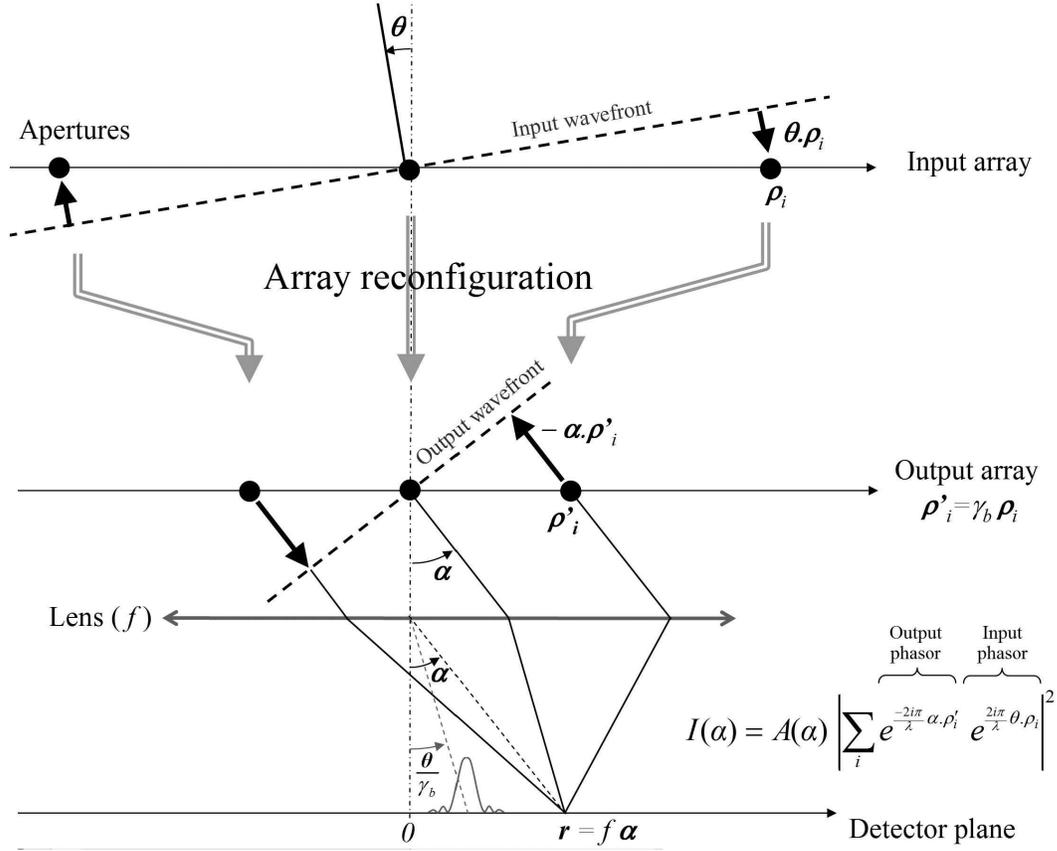}
\hspace*{\fill} \caption{ Optical path differences (OPD) induced
by an array reconfiguration. Compared to the central beam
($\rho=\rho'=0$), the OPD cumulated by the $i^{th}$ beam from the
entrance sub-pupil (position $\rho_i$) until the exit sub-pupil
(position $\rho'_i$) is $\bm{\theta}\cdot\bm{\rho_i}$ (entrance
OPD due to the source off-axis) minus
$\bm{\alpha}\cdot\bm{\rho'_i}$ (exit OPD due to the array
reconfiguration), with $\bm{\alpha}$ the angle position considered
on the image plane. If the array reconfiguration is an homothetic
remapping such as $\bm{\rho'_i}=\gamma_b\,\bm{\rho_i}$, a central
peak is formed on the detector at the angular position
$\bm{\theta}/\gamma_b$. The instrumental OPD is assumed to be null
(\ie the interferometer is cophased on-axis).}
\label{fig:formal_array}
\end{figure*}

To provide a directly exploitable image, a combiner may
\apriori perform a transformation of the wavefront at two
different spatial frequency scales:

\begin{itemize}
\item a ``high-frequency transformation'', which consists in
displacing the sub-apertures centers, therefore altering the
geometry of the array, \item a ``low-frequency transformation'',
which affects only the beams of each sub-aperture individually,
such as a beam compression, a beam deflection, a pupil plane
conjugation, a spatial filtering, etc.
\end{itemize}

The array is now reconfigured: let $\bm{\rho_i}$ and
$\bm{\rho'_{i}}$ respectively represent the position vectors of
the $i^{th}$ aperture in the entrance and output pupil planes. Let
also be $\bm{\theta}$, the off-axis of the source, and
$\bm{\alpha}$ the position vector in the image plane. Assuming
that the interferometer is cophased on-axis (\ie the optical paths
of all arms are equal), the total optical path of the $i^{th}$
beam is $\bm{\theta}\cdot\bm{\rho_{i}}-\bm{\alpha} \cdot
\bm{\rho'_{i}}$ (Fig. \ref{fig:formal_array}). Because of the
additional term induced by the remapping, the PSF becomes non
translation invariant: we have to update equation \ref{equ:gen0}
that is now $\bm{\theta}$-dependant:
\begin{eqnarray}
I(\bm{\alpha},\bm{\theta})& = & A(\bm{\alpha}) \nonumber
\\
&\times& \bigg|\sum_{i=1}^{n_T} \exp \frac{-2i\pi\: \bm{\alpha}
\cdot \bm{\rho'_{i}}}{\lambda} \cdot \exp \frac{2i\pi\:
\bm{\theta}\cdot\bm{\rho_{i}}}{\lambda}\bigg|^2.
\label{equ:gen_array}
\end{eqnarray}

The first exponential term is the fringe pattern which depends
only on the output pupil arrangement ($\bm{\rho'_{i}}$). The
second exponential term contains the piston induced by the source
off-axis $\bm{\theta}$.

If the reconfiguration of the array is homothetic, we can
introduce $\gamma_b$, so that $\bm{\rho'_i}=\gamma_b \:
\bm{\rho_i}$. Equation \ref{equ:gen_array} becomes:
\begin{equation}
I(\bm{\alpha},\bm{\theta}) = A(\bm{\alpha}) \times
\bigg|\sum_{i=1}^{n_T} \exp \frac{-2i\pi (\bm{\alpha}-
\bm{\theta}/\gamma_b) \cdot \bm{\rho'_{i}}}{\lambda}\bigg|^2.
\label{equ:homo}
\end{equation}

Now, if we denote as $O(\bm{\theta})$ the object intensity
distribution and
\begin{equation}
I_0(\bm{\alpha})=\bigg|\sum_{i=1}^{n_T} \exp \frac{-2i\pi\:
\bm{\alpha} \cdot \bm{\rho'_{i}}}{\lambda}\bigg|^2 \label{equ:Io}
\end{equation}
\noindent the interference function, we can write, from equation
\ref{equ:homo}, the intensity distribution for an extended source:
\begin{equation}
I(\bm{\alpha})=\int\!\!\!\int O(\bm{\theta}) A(\bm{\alpha}) \:
I_0(\bm{\alpha}- \bm{\theta}/\gamma_b ) \:d^2\bm{\theta}.
\label{equ:integral}
\end{equation}

Assuming that envelope $A$ is fixed and its extent is larger than
the object size, one can take $A$ out of the integral and
approximate equation \ref{equ:integral} as a convolution product
(the normalization factor has been eliminated for readability):

\begin{equation}
I(\bm{\alpha})\approx A(\bm{\alpha}) \cdot
O(\gamma_b\,\bm{\alpha}) \otimes I_0(\bm{\alpha})\,.
\label{equ:pseudo_conv}
\end{equation}

A convolution relationship between the object and the image
remains inside the envelope, provided that the beam combiner keeps
the pattern of the sub-aperture centers unchanged. This condition
is less restrictive than the original formulation of the golden
rule \citep{traub86} claiming that a strict homothetic mapping,
including the sub-pupils (\ie the Fizeau scheme), was required.
This rule is true only if an infinite DIF is required. But, as we
have demonstrated in section \ref{sec:nt}, an infinite DIF is
useless for a diluted interferometer since the exploitable field
is limited by the incomplete coverage of the $(u,v)$ plane.

If an homothetic reconfiguration of the array preserves the
convolution relation over a finite FOV, a non-homothetic one
completely destroys it. Such a remapping however may be useful to
make the sparse interferometric pupil suitable for coronagraphy.
In this particular case, a second remapping restores the geometry
of the array, after the coronagraph, to recover the convolution
relationship \citep{guyon02}.

Equation \ref{equ:pseudo_conv} has been deduced using the
hypothesis of a quite wide, immobile envelope position: the
following subsections detail what happens to the envelope with
real combiners.

\begin{figure*}
\hfill\includegraphics[width=.9\textwidth]{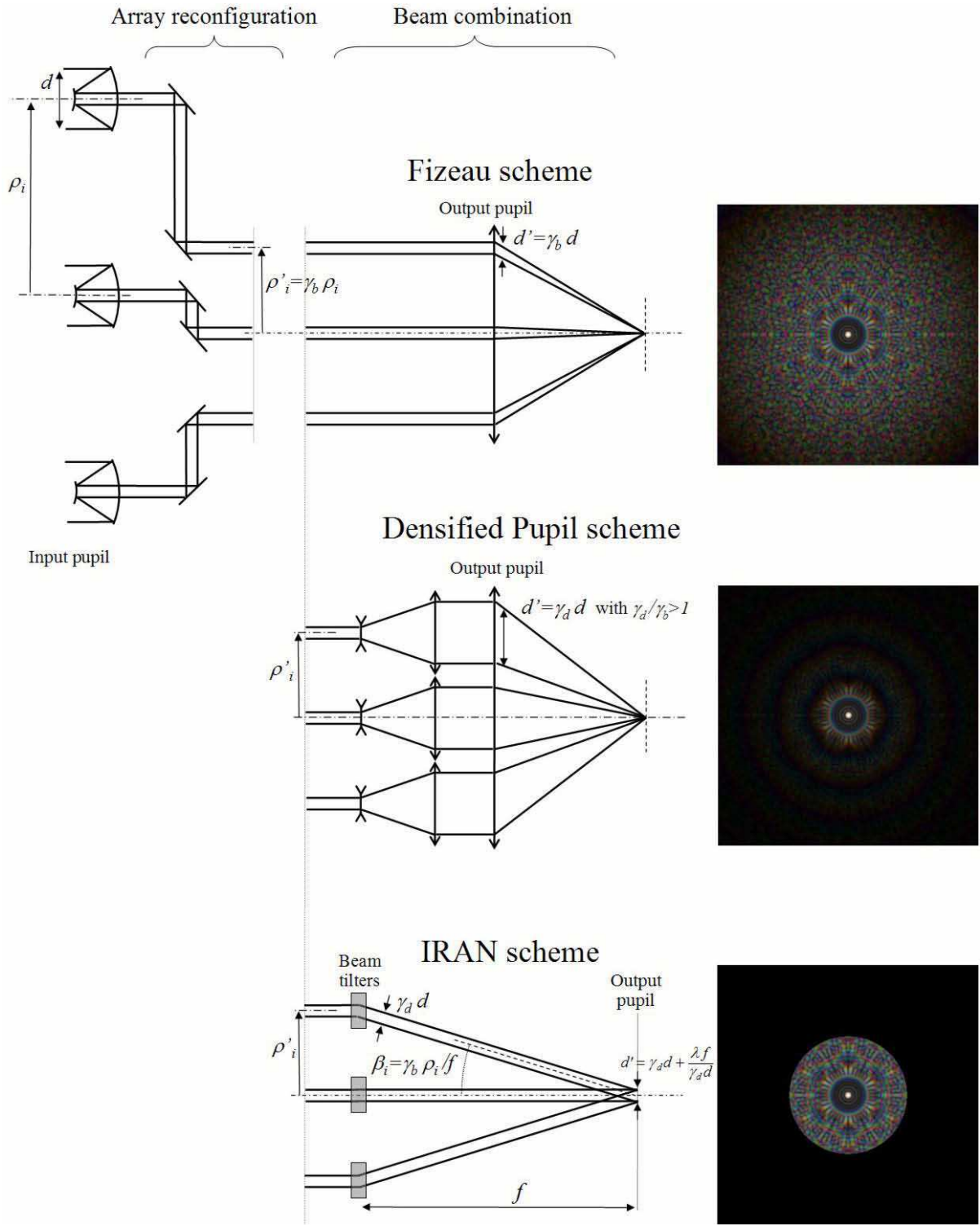}
\hspace*{\fill} \caption{The three beam combination schemes
considered for direct imaging (Fizeau, Densified Pupil and IRAN)
and their corresponding PSF for a KEOPS-like array. A very generic
version of IRAN, called IRAN$^b$ by \citet{aristidi}, is
represented here. A previous version, called IRAN$^a$
\citep{vakili04}, forms an intermediate image plane. Both versions
of IRAN are strictly equivalent and are described by the same
formalism. The Fizeau PSF spreads in numerous dispersed sidelobes
over $\lambda/d$, while the Densified Pupil and the IRAN schemes
can concentrate more flux inside the usable clean field (CLF)
thanks to a narrower envelope (an Airy-shaped envelope and a flat
top-hat envelope respectively). Relay lenses forming the output
pupil have been omitted for clarity, and the beam diffraction is
ignored for the IRAN PSF calculation (polychromatic PSF:
$\Delta\lambda/\lambda=0.2$, intensity scale in power 0.3).}
\label{fig:bsc}
\end{figure*}

\subsection{Fizeau combination}
\label{sec:fiz}

The first all-to-one beam combination scheme proposed for
direct imaging was a strict homothetic mapping scheme, also
called Fizeau, where the output pupil (seen from the focal plane)
is a reduced copy of the entrance pupil (seen from the sky).

For the Fizeau scheme, the homothety ratio of the baselines
$\gamma_b$ is compensated by an identical homothety of the
sub-apertures $\gamma_d = d'/d$, where $d$ and $d'$ represent the
diameter of a sub-aperture in the entrance and in the output pupil
plane respectively (Fig. \ref{fig:bsc}). In this way, envelope $A$
becomes the diffraction lobe of an aperture of diameter
$\gamma_d\, d$, referred to as $A_{\gamma_d\, d}$. The beam
compression increases the slope of the wavefront of each beam by
the factor $1/\gamma_d$, therefore shifting the envelope to the
angular position $\bm{\theta}/\gamma_d$ (with $\bm{\theta}$ the
source off-axis). Equation \ref{equ:homo} can now be rewritten as:
\begin{eqnarray}
I(\bm{\alpha},\bm{\theta}) &=& A_{\gamma_d\,d}\left(\bm{\alpha}-
\frac{\bm{\theta}}{\gamma_d}\right) \nonumber
\\
&\times& \bigg|\sum_{i=1}^{n_T}
\exp{\frac{-2i\pi}{\lambda}\bigg(\bm{\alpha}-\frac{\bm{\theta}}{\gamma_b}\bigg)
\cdot \bm{\rho'_{i}}}\bigg|^2. \label{equ:fiz_pd}
\end{eqnarray}

Since $\gamma_d=\gamma_b$, interference pattern and envelope are
translated of the same amount ($\bm{\theta}/\gamma_b$): the Fizeau
PSF is translation invariant (Fig. \ref{fig:psf-off-axis}) and the
image of an extended source can be rigorously written as a
convolution product:
\begin{equation}
I(\bm{\alpha})= O(\gamma_b\,\bm{\alpha}) \otimes
\big[A_{\gamma_b\,d}(\bm{\alpha})\cdot I_0(\bm{\alpha})\big].
\label{equ:conv:fiz}
\end{equation}

\subsection{Densified Pupil combination}
\label{sec:dp}

The technique of pupil densification was introduced
\citep{labey96} to provide more luminous images than in the Fizeau
case: it concentrates the extremely wide and dispersed Fizeau PSF
into the central interference peak (Fig. \ref{fig:bsc}). The
sparse aperture of an interferometer is now compatible even with
coronagraphy \citep{labey96,guyon02,riaud}. Thanks to pupil
densification, the interferometer can work very much like any
conventional telescope. This distinctive use of an interferometer
has been called ``hypertelescope''.

According to Labeyrie, a hypertelescope is a multi-aperture
interferometer where the detecting camera is illuminated through
an exit pupil which is a densified copy of the entrance aperture.
``Densified copy'' implies that the pattern of exit pupil centers
is conserved with respect to the entrance pattern, while the size
of the elementary sub-pupils is magnified, for example by an
inverted Galilean telescope (Fig. \ref{fig:bsc}). The
densification is quantified by a dimensionless,
convention-independent coefficient $\gamma = \gamma_d/\gamma_b$.
This technique is a simple 2-D generalization of Michelson's
original truss \citep{michelson}.

When the pupil is densified, equation \ref{equ:fiz_pd} remains
valid but this time, $\gamma_d/\gamma_b\!\!>\!\!1$. For an
off-axis source (position $\bm{\theta}$), the fringe pattern is
shifted to the angular position $\bm{\theta}/\gamma_b$, while the
envelope is shifted to a lower angular position:
$\bm{\theta}/\gamma_d$ (Fig. \ref{fig:psf-off-axis}). There is
therefore no more strict object-image convolution relationship and
equation \ref{equ:integral} has to be updated. The densified pupil
image is given by:
\begin{equation}
I(\bm{\alpha})=\!\!\!\int\!\!\!\!\int\! O(\bm{\theta})
\:A_{\gamma_d\,d}\!\bigg(\!\bm{\alpha}\!-
\!\frac{\bm{\theta}}{\gamma_d}\!\bigg) \,
I_0\bigg(\!\bm{\alpha}\!-\!
\frac{\bm{\theta}}{\gamma_b}\!\bigg)\,d^2\bm{\theta}.
\label{equ:triple_conv}
\end{equation}

For highly diluted arrays, the pupil densification can be so
strong (\ie $\gamma_d/\gamma_b\gg1$) that the shift of the
envelope becomes negligible in comparison with the shift of the
fringe pattern. We therefore meet the conditions of validity of
the general equation \ref{equ:pseudo_conv}, and can apply here the
same steps used from Equ. \ref{equ:integral} to Equ.
\ref{equ:pseudo_conv}:
\begin{equation}
I(\bm{\alpha})\approx A_{\gamma_d\,d}(\bm{\alpha}) \cdot
O(\gamma_b\,\bm{\alpha}) \otimes I_0(\bm{\alpha})\,.
\label{equ:conv:pd}
\end{equation}

A convolution relation remains, but in a FOV limited by the
envelope of a densified sub-aperture only.

\subsection{IRAN combination}

The Interferometric Remapping Array Nulling (hereafter IRAN) beam
combination scheme was introduced by \citet{vakili04}, and
proposed for the already mentioned KEOPS project. IRAN was
proposed as an alternative to the hypertelescope reconfiguration
to prevent the loss of classical object-image convolution relation
that it suffers during the pupil remapping. The solution proposed
by the authors consists in recording the interference in a pupil
plane rather than in the image plane (Fig. \ref{fig:bsc}).

This solution exhibits appealing features for direct imaging,
compared to the pupil densification. The moving Airy-shaped
envelope of the hypertelescope is indeed replaced by a flat
top-hat envelope, whose position is independent of the position of
the sources. Thanks to this unique property, the formalism of
image formation in the pupil-plane is simplified (Equ.
\ref{equ:conv:iran}) and the PSF becomes translation-invariant
inside the top-hat.

The IRAN combiner performs again a non purely homothetic remapping
of the wavefront, characterized by the two magnification
coefficients $\gamma_d$ and $\gamma_b$ introduced in the previous
section. Because the IRAN scheme preserves the geometry of the
array, the interference function is unchanged. The image can be
expressed exactly as a convolution product from equation
\ref{equ:pseudo_conv}:
\begin{equation}
I(\bm{\alpha})= P_{d'}(\bm{\alpha}) \cdot O(\gamma_b\,\bm{\alpha})
\otimes I_0(\bm{\alpha})\,, \label{equ:conv:iran}
\end{equation}
\noindent with $P_{d'}$ a top-hat shaped envelope of diameter
$d'$, whose position is independent from the source off-axis (Fig.
\ref{fig:psf-off-axis}), as in the case of a strong pupil
densification. To be rigorous, the diffraction of the output
collimated beams have to be considered, especially if output pupil
size $d'$ becomes comparable to the diffraction lobe (Sec.
\ref{sec:iran:opt}). In this case, the function $P_{d'}$ is more
exactly a top-hat function convolved with an Airy function. The
total width of $P_{d'}$ becomes:
\begin{equation}
d'=\gamma_d\,d+\frac{\lambda\,f}{\gamma_d\,d}\:\:.
\label{equ:d':iran}
\end{equation}

Then, the IRAN envelope is actually flat only in the central part
(the geometrical diameter of the output pupil) and features
diffracted edges (the second term of the sum).

\subsection{Single-mode fiber combination}
\label{sec:fiber}

Compared to classical bulk optics, single-mode fibers offer a very
convenient solution for the beam transportation from the telescope
focus towards the combiner \citep{perrin}. They apply also a
spatial filtering which could improve the performances drastically
in presence of residual phase errors \citep{coude}. This spatial
filtering induces no information loss with highly diluted
interferometers, since we are only interested in the high spatial
frequencies measured by the baselines, and not in low spatial
frequencies measured by one sub-aperture.

Fibers are also of interest in direct imaging for the beam
remapping and densification. Such a combiner has been proposed for
a densified pupil imager on the VLTI \citep{lardi05} and a
demonstrator is under development in laboratory \citep{sirius}.

As the tilt is lost in a single-mode fiber, the fringe envelope
remains on-axis whatever the combination scheme and the
densification factors. Then, the perfect convolution relationship
provided by the Fizeau mode is destroyed by fibers. However, the
Densified Pupil or the IRAN scheme can be used indifferently for
combining the beams at the fiber outputs, the difference being
only in the envelope shape. The fibred versions of both
combination schemes are illustrated and studied in details by
\citet{patru06}. In this paper, we will focus our attention on the
envelope whose width determines directly the DIF.

The amplitude distribution of the output beam is a gaussian law,
generally truncated by a collimating lens of diameter $d'$ and
focal length $f'$. We refer to this distribution as
$\psi_{d'\!\!,f'}(\bm{x})$. Although the notion of pupil vanishes
with fibers, $d'$ defines a new output pupil diameter, and
$\gamma_d$ can again be defined as the ratio $d'/d$.

\subsubsection{Fibred Densified Pupil scheme}

As the Densified Pupil scheme forms the fringes in an image plane,
the intensity distribution of the image is:
\begin{equation}
I(\bm{\alpha})=|\mathrm{FT}(\psi_{d'\!\!,f'}(\bm{x}))|^2\times
I_0(\bm{\alpha}-\bm{\theta}/\gamma_b), \label{equ:fib:dp:I}
\end{equation}
\noindent where FT denotes a Fourier Transform and $I_0$ is the
interference function defined in equation \ref{equ:Io}.  Thus, the
envelope becomes $|\mathrm{FT}(\psi_{d'\!\!,f'}(\bm{x}))|^2$ and
is not a pure Airy function as with classical optics (sec.
\ref{sec:dp}) but a gaussian lobe convolved with an Airy lobe. As
the envelope remains on-axis, we meet the exact conditions of
validity of the general equation \ref{equ:pseudo_conv}. Then, from
equation \ref{equ:fib:dp:I}, we can write the image of an extended
object $O$ as a convolution relationship:
\begin{equation}
I(\bm{\alpha})=|\mathrm{FT}(\psi_{d'\!\!,f'}(\bm{x}))|^2\cdot
O(\gamma_b\,\bm{\alpha})\otimes I_0(\bm{\alpha}).
\label{equ:fib:dp:conv}
\end{equation}

Because the output beams have a gaussian profile, a full pupil
densification is impossible. Then, there is a trade-off between
the sensitivity gain provided by the densification and the flux
lost by the beam truncation. \citet{patru06} report that the
maximum intensity is reached for $d'=2.2\,\omega(f')$, with
$\omega(f')$ the radius where the amplitude is $1/e$ times the
maximum amplitude.

\subsubsection{Fibred IRAN scheme}

The IRAN scheme can also be used for combining beams behind
single-mode fibers \citep{patru06}. If the diffraction is
negligible, the intensity distribution of the image is:
\begin{equation}
I(\bm{\alpha})=|\psi_{d'\!\!,f'}(\bm{x})|^2\times
I_0(\bm{\alpha}-\bm{\theta}/\gamma_b),
\end{equation}
\noindent meaning that the envelope is directly the truncated
gaussian profile and not its Fourier Transform as for the
Densified Pupil case. Once again, as the envelope position is
independent from $\bm{\theta}$, we can write the image of an
extended object $O$ as a convolution product:
\begin{equation}
I(\bm{\alpha})=|\psi_{d'\!\!,f'}(\bm{x})|^2\cdot
O(\gamma_b\,\bm{\alpha})\otimes I_0(\bm{\alpha}).
\label{equ:fib:iran:conv}
\end{equation}

As the maximum of the function $|\psi_{d'\!\!,f'}(\bm{x})|^2$ is
always 1 whatever $d'$, the beam truncation has no effect on the
central peak intensity and reduces the field only by vignetting.

\begin{figure*}
\hfill\includegraphics[width=.9\textwidth]{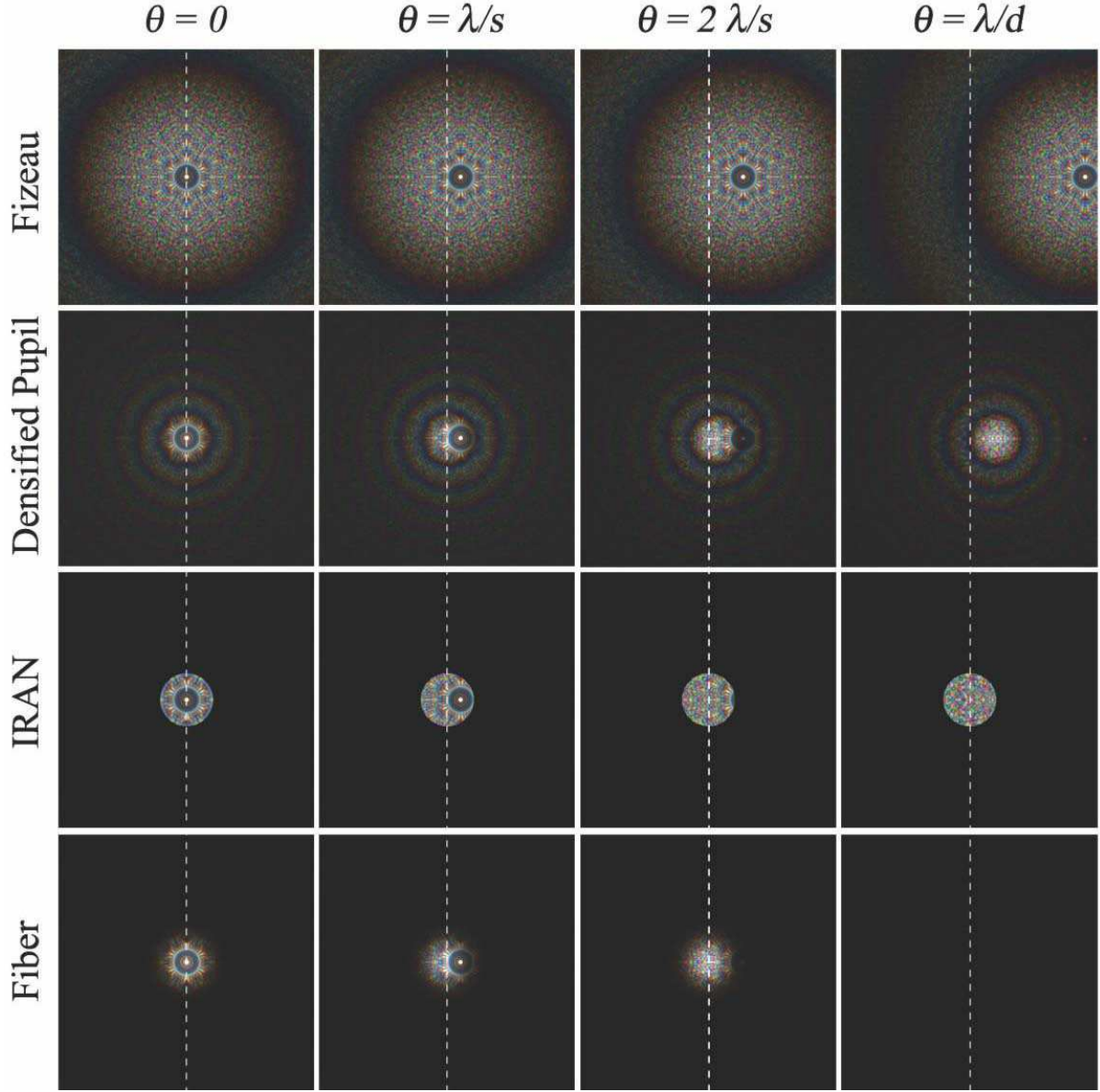}
\hspace*{\fill} \caption{Fizeau PSF for different source off-axis,
compared to PSF provided by a Densified Pupil, an IRAN and a
fibred combination scheme (partial densification:
$\mathrm{DIF}=2\,\lambda/s$). The $\theta=0$ and
$\theta=\lambda/s$ columns show a source inside the DIF for all
the schemes. The $\theta=2\lambda/s$ column illustrates a source
outside the DIF for all the schemes, excepted for the Fizeau case.
The last column shows a source at the edge of the coupled field
(CF) for the Fizeau, Densified Pupil and fibred schemes, but
inside the CF for the IRAN scheme (polychromatic PSF:
$\Delta\lambda/\lambda=0.2$, intensity scale in power $0.3$, beam
diffraction is ignored for IRAN).} \label{fig:psf-off-axis}
\end{figure*}

\section{FOV comparison and optimization}
\label{sec:comp}

The previous section has shown that the choice of the beam
combination scheme does not affect the interference function $I_0$
if the array pattern is unchanged. The beam combiner simply
applies, in a way, a ``windowing'' on this function, concentrating
more flux on the central part of $I_0$. The width of this
``windowing'' defines the DIF. This section will compare the DIF
of each beam combiner and will demonstrate that the Densified
Pupil and the IRAN schemes can be optimized in order to match the
DIF to the useful FOV, provided by the (u,v) plane coverage, \ie
the CLF.

\subsection{DIF comparison}

\subsubsection{The Fizeau DIF}

In a Fizeau scheme, an interferometer works strictly like a giant
masked telescope magnifying the object $1/\gamma_b$ times and
preserving the convolution relationship between the object and the
image over an infinite FOV (Equ. \ref{equ:conv:fiz}). The
practical limits of the DIF are then imposed only by geometrical
aberrations, vignetting or by the size of the detector. This
advantage will be discussed in section \ref{sec:dis}.

\subsubsection{The Densified Pupil DIF}

Unlike the Fizeau case, the interference function is now modulated
by an envelope of small diameter, which drastically reduces the
DIF (Equ. \ref{equ:conv:pd}). A non resolved off-axis source
(position $\bm{\theta_0}$) can \apriori be correctly imaged as
long as its central interference peak remains within the envelope
(inside the FWHM by convention), \ie:
\begin{equation}
\frac{\bm{\theta_0}}{\gamma_b}-\frac{\bm{\theta_0}}{\gamma_d}
\le\frac{\lambda}{2\,\gamma_d\,d}\:\:,
\end{equation}

\noindent which constraints the diameter of the DIF, in radians on
the sky:
\begin{equation}
DIF_{\oslash}\approx \frac{\lambda}{(\gamma_d/\gamma_b-1)\:d}\:\:.
\label{equ:dif}
\end{equation}

For the fibred version of the Densified Pupil scheme, the envelope
remains on-axis (Equ. \ref{equ:fib:dp:conv}). Then the DIF
directly corresponds to the width of the envelope
$|\mathrm{FT}(\psi_{d'\!\!,f'}(\bm{x}))|^2$ (the FWHM by
convention). For the optimal gaussian beam truncation, we find:
\begin{equation}
\mathrm{DIF_\oslash}\approx1.13\;\frac{\gamma_b}{\gamma_d}\cdot
\frac{\lambda}{d}\:\,.
 \end{equation}

\subsubsection{The IRAN DIF}

The DIF of the IRAN scheme corresponds to the sky angular extent
imaged over the output beams diameter $d'$. Considering that
$\gamma_b$ is the angular scaling factor from the sky to the
detector and that the angular size of the output pupil, seen from
the beam tilters, is $d'/f$ (Fig. \ref{fig:bsc}), the DIF size on
the sky can be written in radians from equation \ref{equ:d':iran}
as:
\begin{equation}
\mathrm{DIF_\oslash}=\frac{\gamma_b\,\gamma_d\,d}{f}+\frac{\gamma_b\,\lambda}{\gamma_d\,d}\:.
\label{equ:iran:dif:rad}
\end{equation}

For the fibred version of the IRAN scheme, the output beam extent
is now limited by the diameter of the collimating lens or by the
detector size. If the output beam diameter is written
$\gamma_d\,d$, then equation \ref{equ:iran:dif:rad} remains valid
for the fiber case.

\subsection{Optimization of the DIF}

The reduction of the DIF due to the pupil densification, compared
to the Fizeau, is not at all a drawback as one could believe at
first sight. The densification coefficients $\gamma_b$ and
$\gamma_d$ are free parameters that can be used to concentrate
more flux inside the field of interest. For instance, the
densification can be adjusted to match the DIF to the CLF in order
to maximize the luminosity gain. An analogue FOV optimization can
\apriori be found for the IRAN scheme.

\subsubsection{Densified Pupil scheme optimization }

The optimal pupil densification factors can be found by equalizing
the equations \ref{equ:clf} and \ref{equ:dif}. Assuming
$\gamma_d\gg\gamma_b$, the pupil densification factor have to meet
the following condition:
\begin{equation}
\frac{\gamma_d}{\gamma_b}\approx\frac{s}{d}\:\:,
\label{equ:pd:dif=clf}
\end{equation}

\noindent with $s$ the smallest typical spacing between apertures,
and $d$ the aperture diameter. This optimum occurs when two
adjacent apertures of the array have their output pupils in
contact. Note that this pupil densification was precisely the one
originally proposed by \citet{labey96}, but without introducing
FOV considerations.

\subsubsection{IRAN scheme optimization}
\label{sec:iran:opt}

As for the hypertelescope mode, we can determine the densification
factors $\gamma_d$ and $\gamma_b$ equalizing the DIF on the CLF
for the IRAN combination scheme. To keep the main advantage of the
IRAN, which is the flatness of the envelope $P_{d'}$, we choose to
equalize only the flat part of the DIF (the first term of equation
\ref{equ:iran:dif:rad}) with the CLF. From equations
\ref{equ:iran:dif:rad} and \ref{equ:clf}, this implies:
\begin{equation}
\gamma_d\:\gamma_b=\frac{\lambda\:f}{s\:d}\:\,.
\label{equ:iran:dif=clf}
\end{equation}

This equation highlights the fact that the optimum is reached only
for one wavelength, except if at least one of the densification
factors ($\gamma_b$ or $\gamma_d$) is proportional to $\lambda$.
This chromatic issue does not exist in the Densified Pupil case.
Indeed, if $\gamma_d$ and $\gamma_d$ are achromatic, the width of
the envelope $P_{d'}$ is not purely proportional to $\lambda$
(Equ. \ref{equ:d':iran}), and cannot exactly follow the natural
chromaticity of the CLF extent ($\lambda/s$).

By introducing $s'=\gamma_b\,s$, the distance separating two
adjacent afocal beams after the array remapping, equation
\ref{equ:iran:dif=clf} becomes:
\begin{equation}
s'=\frac{\lambda}{\gamma_d\,d}\,f\:, \label{equ:iran:s'}
\end{equation}

\noindent meaning that the output beams are only separated by a
diffraction lobe. In case of the IRAN$^a$ scheme (\cf caption of
figure \ref{fig:bsc}), this condition occurs when the center of a
sub-image coincides with the first zero of the neighboring
sub-image. In other words, the intermediate image plane has to be
fully densified. This result reminds us of the analogy with the
Densified Pupil scheme: there is only an exchange between the
pupil and the image planes.

If classical bulk optics are used, the minimal practical value of
$s'$ is reached when the output afocal beams are in contact, \ie
$s'=\gamma_d\,d$. Then equation \ref{equ:iran:s'} gives  a new
condition for $\gamma_b$ and $\gamma_d$:
\begin{equation}
\frac{\gamma_d}{\gamma_b}=\frac{s}{d}\:\,.
\label{equ:iran:ratio_gam}
\end{equation}

Thanks to equations \ref{equ:iran:dif=clf} and
\ref{equ:iran:ratio_gam} it is possible to find a unique solution
for both densification factors:
\begin{eqnarray}
\gamma_b=\frac{\sqrt{\lambda\,f}}{s} &,&
\gamma_d=\frac{\sqrt{\lambda\,f}}{d}\:. \label{equ:iran:factors}
\end{eqnarray}

With these values, the central flat part of the DIF is equal to
the diffractive part (Equ. \ref{equ:iran:dif:rad}). Then the total
DIF is twice wider than the CLF (\ie $2\,\lambda/s$).

\subsection{Luminosity gains}
\subsubsection{Densified Pupil luminosity gain}

The reason to be of pupil densification is to concentrate useless
sidelobes flux in the central peak in order to provide much more
luminous images than what the Fizeau allows. Actually, the pupil
densification shrinks the carrier envelope $A$ by a factor
$\gamma_d/\gamma_b$ in comparison to the Fizeau mode. Because the
energy is conserved, the image is much brighter than in the Fizeau
case, the gain being:
\begin{equation}
G=\bigg(\frac{\gamma_d}{\gamma_b}\bigg)^2. \label{equ:pd:intens}
\end{equation}

In the optimized case, where the DIF and the CLF coincide, the
intensity gain reaches its maximum. From equations
\ref{equ:pd:dif=clf} and \ref{equ:pd:intens}, the best possible
sensitivity gain is :
\begin{equation}
G_0=\bigg(\frac {s}{d}\bigg)^2. \label{equ:pd:intens:max}
\end{equation}

If single-mode fibers are used, the luminosity gain remains
unchanged because the envelope width is almost unchanged. The only
significant difference compared to bulk optics is the flux loss
induced by the beam injection into the fiber ($22\%$ in the best
case) and by the output beam truncation ($8\%$ in the optimized
case \citep{patru06}).

\subsubsection{The IRAN luminosity gain}

No considerations about the sensitivity were made by the authors
of the IRAN scheme and yet, IRAN also can bring a valuable gain in
luminosity compared to the Fizeau mode. Indeed, there is an
intensification of the central peak if the support $P_{d'}$ of the
fringe pattern is narrower than $\lambda/d$, the \emph{equivalent
width} of the Fizeau envelope. From equation
\ref{equ:iran:dif:rad}, the intensity gain provided by the IRAN
scheme compared to the Fizeau is:

\begin{equation}
G=\bigg(\frac{\gamma_d}{\gamma_b}\cdot\frac{\lambda\:f}{\gamma_d^2
\:d^2+\lambda\,f}\bigg)^2. \label{equ:iran:intens}
\end{equation}

The best possible intensity gain, keeping the advantage of IRAN
(the flat envelope), is reached when the flat part of the DIF
coincides with the CLF. Its value is deduced from equations
\ref{equ:iran:factors} and \ref{equ:iran:intens}:
\begin{equation}
G_0\approx\bigg(\frac{s}{2\,d}\bigg)^2,\label{equ:iran:intens:max}
\end{equation}

\noindent that is 4 times lower than the optimal Densified Pupil
case (Equ. \ref{equ:pd:intens:max}).

\subsection{Coupled FOV (CF)}
\subsubsection{Definition}
A beam combiner does not differ from others only in its DIF
extent, but also in its coupled FOV (CF) extent. By analogy with
spatial filters and optical fibers, we say that a source is
``coupled'' to the interferometer (\ie located inside the CF), if
some of its photons appear on-axis in the detector plane.

The CF is an important parameter to take into consideration
because it determines the ease with which the interferometer may
reach the crowding limit (Sec. \ref{sec:nt}). Indeed, the image of
an on-axis object is polluted by the sidelobes of all surrounding
sources present in the CF.

\subsubsection{Coupled FOV comparison}
By convention, we assume that a source belongs to the CF if the
on-axis pixel is located within the envelope $A$ of the source.
For the Fizeau case, the CF extent is $\lambda/d$, \ie the radius
of the diffraction lobe of a sub-aperture (Fig.
\ref{fig:psf-off-axis}).

For the Densified Pupil case, the CF can be determined from
equation \ref{equ:fiz_pd}. The edge of the envelope
$A_{\gamma_d\,d}$ coincides with the central pixel when the source
angular separation $\bm{\theta_0}$ meets the condition:

\begin{equation}
\frac{\bm{\theta_0}}{\gamma_d} = \frac{\lambda}{\gamma_d\,d}\:\:.
\end{equation}

Then, the CF is not dependent on the pupil densification factor
and it is equal to $\lambda/d$, exactly as in the Fizeau case: the
shrinking of the envelope is indeed compensated by the slower
shift of the envelope (Fig. \ref{fig:psf-off-axis}).

For IRAN, the fringe pattern is recorded in a pupil plane. The
corresponding CF is therefore infinite by definition (Fig.
\ref{fig:psf-off-axis}), and the crowding becomes a serious issue.
The use of a spatial filtering (pinhole or single mode fiber in
each sub-image plane) can overcome this limit, but such a spatial
filtering smoothes the output pupil edges. If a hole picks up only
$\lambda/d$ in each beam, the envelope $P_{d'}$ becomes an Airy
function exactly as for the Fizeau or the Densified Pupil schemes,
and IRAN loses its main advantage.

Lastly, if single-mode fibers are used for the beam combination,
the CF is obviously limited to $\lambda/d$ (Fig.
\ref{fig:psf-off-axis}) whatever the chosen scheme (Densified
Pupil or IRAN).

\begin{table}
\hfill
\renewcommand{\arraystretch}{1.7}
\begin{tabular}{|c | c | c |}
   \hline
   & DIF & CF \\
   \hline
   Fizeau & $\infty$ & $\lambda/d$ \\
   Densified Pupil & ${\displaystyle\frac{\lambda}{(\gamma_d/\gamma_b-1)\,d}}$ & $\lambda/d$ \\
   IRAN & $\gamma_b\,\gamma_d\;d/f$  & $\infty$ \\
   Fibred DP & $\gamma_b/\gamma_d\cdot\lambda/d$  & $\lambda/d$ \\
   Fibred IRAN & $\gamma_b\,\gamma_d\,d/f$ & $\lambda/d$ \\
   \hline
\end{tabular}
\renewcommand{\arraystretch}{1.}
\hspace*{\fill} \caption{ Direct Imaging Field (DIF) and Coupled
Field (CF) of different combination schemes (in radians on the
sky). The Fizeau DIF and the IRAN CF may formally be infinite, but
some practical constraints, such as geometrical abberations,
vignetting or detector size, will limit these fields.}
\label{tbl:fov}
\end{table}

\section{Discussions} \label{sec:dis}

\subsection{Identification of different FOV limitations}

The keystone of this discussion about direct imaging with an
interferometer is definitely the notion of field of view. It is a
delicate notion that requires the introduction of four possible
limitations:
\begin{itemize}
\item the information FOV (IF), which is the maximum angular size
that a compact source can have without exceeding the crowding
limit,

\item the clean FOV (CLF), which corresponds to the useful clean
central part of the PSF,

\item the direct imaging FOV (DIF), inside which a source can be
directly imaged by the interferometer, and

\item the coupled FOV (CF), inside which any source spreads
photons on the detector (for better or for worse).
\end{itemize}

The IF and the CLF are defined by the array configuration
(respectively by the number of unique baselines and by the
smallest typical baseline), while the DIF and the CF are imposed
by the beam combination scheme. In addition to these four
different fields, we can mention the existence of a fifth one: the
``astrometric'' FOV, over which high resolution measurements of
angular separations are possible. As this field is generally not
contiguous, we choose not to talk about wide field interferometry,
contrarily to some authors \citep{montilla} and refer to it as
multi-field observation mode since it is an extension of the
dual-field mode using differential delay-lines.

\subsection{Choice of the beam combination scheme}

Despite the unique characteristic of an infinite DIF, the use of a
Fizeau combiner for highly diluted apertures appears excluded,
because of PSF quality considerations. Indeed, the Fizeau PSF
spreads in numerous secondary spectrally dispersed peaks or
speckles over a wide area (the diffraction lobe of a
sub-aperture), which makes it poorly suited to high dynamic-range
imaging (Fig. \ref{fig:bsc} and \ref{fig:psf-off-axis}). This is
the reason that motivated the development of other beam combiners
such as the Densified Pupil and IRAN.

Moreover, a large DIF does not mean that any extended sources can
be imaged properly. Section \ref{sec:archi} has proved that the
ultimate wide field imaging capabilities of an interferometer are
already limited by the aperture array configuration (hence by the
(u, v) plane coverage). Indeed, the finite number of apertures
limits the maximum quantity of information recordable in a
snapshot, while the shortest baseline $s$ determines the angular
size of the usable clean field (CLF).

Despite these considerations, the Fizeau scheme remains of
interest for observing in a multi-field mode. The Fizeau focus
acts as a natural spatial filter, which means that two sources
separated by more than $\lambda/d$ (the CF) can be observed
simultaneously without interacting with each other. Such an
observing mode preserves astrometric precision and involves
neither star separators nor extra differential delay lines,
contrarily to conventional dual field mode imaging.

A very attractive configuration therefore is to put an array of
pupil densifiers and/or IRAN modules after the Fizeau focus. Such
multi-field scheme benefits from both techniques. The Fizeau
provides a wide FOV but unusable images, while the pupil
densifiers focus on smaller fractions of the field and provide
simultaneous directly exploitable images, as proposed in
\citet{labey03}. With this combination, all high angular
resolution techniques used with conventional telescopes can be
employed: speckle imaging with partial AO corrections or long
exposure imaging, image deconvolution, etc. Each densifier may
also include a coronagraph which opens the doors to high dynamic
imaging with an interferometer.

The discussion is now about the choice between the Densified Pupil
and the IRAN scheme. Our comparative study reveals that both
combination schemes are fundamentally similar, the only difference
being an exchange between the pupil and the image planes. This
difference vanishes if single-mode fibers are used in either
scheme, since the envelope has nearly the same shape and always
remains on-axis (\cf fig. \ref{fig:psf-off-axis} and table
\ref{tbl:fov}).

Both schemes can be optimized to equalize the DIF with the FOV
delivered by the (u, v) plane coverage. This field optimization
also induces an advantageous luminosity gain compared to the
Fizeau mode, which can rise up to $10^4$ for kilometric arrays
(Equ. \ref{equ:pd:intens:max} and \ref{equ:iran:intens:max}). In
the optimized configuration, the hypertelescope mode features a
fully densified output pupil, while IRAN features a fully
densified intermediate image plane.

The classical bulk optics version of IRAN is unique because it
offers a flat DIF, a particularity which can make the image
deconvolution easier \citep{aristidi}. However, its infinite CF
makes IRAN not compatible for observing very extended sources or
rich fields. Moreover, the IRAN scheme cannot concentrate the flux
as efficiently as the hypertelescope does, because of the
diffraction of the output beams. Indeed, equations
\ref{equ:pd:intens:max} and \ref{equ:iran:intens:max} show that
the luminosity gain is 4 times lower than the optimal Densified
Pupil case. To reach the same gain, the size of the IRAN DIF
should be decreased by a factor 2, meaning that the geometrical
diameter of the output pupil should be equal to zero in order to
keep only the diffractive part. This implies that the fringes
should be formed in an image plane rather than in a pupil plane,
exactly as the Densified Pupil scheme does.

Therefore, the argument of an enhanced FOV for the IRAN scheme in
comparison with the hypertelescope mode, initially invoked by
\citet{vakili04}, should be moderated. The previous sections have
already shown that the array itself induces a limit to the FOV
which, anyway, cannot be overcome by the beam combiner.

\section{Conclusion}

The formal analysis introduced here demonstrates that the direct
imaging capabilities of a diluted interferometer, \ie the crowding
limit, the PSF quality and the clean FOV, are determined by the
choice of the geometry of the array only, and not by the choice of
the beam combination scheme. This choice must be motivated by the
science cases which will impose the necessary number of telescopes
and decide to emphasize either FOV or dynamic range.

Among the available beam combination schemes for direct imaging,
Densified Pupil and IRAN have been proven equivalent, except for a
pupil-image plane inversion that changes the shape of a modulating
envelope. The possibility to densify the array actually provides
an optimal image reconstruction technique that fully exploits the
field limited by the $(u,v)$ plane coverage. It drastically
improves the poor quality of the Fizeau PSF by concentrating the
flux on a clean field of view only, while preserving all high
angular resolution information.

Other aspects concerning the direct imaging, such as chromatic
effects of beam combiners, array optimization and coronagraphy,
will be studied in following papers. This study should provide
sufficient matter to define the concept and the instrumentation of
a future direct imaging large array, according to the top-level
requirements (angular and spectral resolutions, FOV,
dynamic-range, bandwidth, etc.) and the primary scientific goals.

\section*{Acknowledgments}

The authors are very grateful to Denis Mourard and Antoine
Labeyrie for fruitful discussions and also to Chris Haniff, the
referee, for his comments on the manuscript.

\end{document}